\documentclass[aps,pre,twocolumn,superscriptaddress,showpacs]{revtex4-1}
\usepackage{epsfig,color,graphics}
\usepackage{siunitx}
\usepackage{amsmath, mathrsfs}
\usepackage{comment}
\usepackage{upgreek}
\usepackage[normalem]{ulem}
\usepackage{float}
\usepackage[latin1]{inputenc}
\renewcommand{\bm}[1]{\boldsymbol{\mathbf{#1}}}
\providecommand*{\pp}[3][]{\frac{\partial^{#1}#2}{\partial #3^{#1}}}
\providecommand*{\dd}[3][]{\frac{\mathrm{d}^{#1}#2}{\mathrm{d} #3^{#1}}}
\providecommand*{\ex}{\bm{e}_{X}}
\providecommand*{\ey}{\bm{e}_{Y}}
\providecommand*{\rmd}{\mathrm{d}}
\newcommand{\qmbox}[1]{\quad \mbox{#1} \quad}
\providecommand*{\ppi[3][]}{\partial_{#3}^{#1}#2}

\begin{document}
\title{Rotation of a submerged finite cylinder moving down a soft incline}
\author{Baudouin Saintyves}
\email{bsy@mit.edu}
\affiliation{Department of Mechanical Engineering, Massachusetts Institute of Technology - Cambridge, MA 02139, USA}
\affiliation{School of Engineering and Applied Sciences, Harvard University, Cambridge, MA 02138, USA}
\author{Bhargav Rallabandi}
\affiliation{Department of Mechanical Engineering, University of California, Riverside, California 92521, USA}
\author{Theo Jules}
\affiliation{School of Engineering and Applied Sciences, Harvard University, Cambridge, MA 02138, USA}
\affiliation{Department de Physique, \'{E}cole Normale Sup\'{e}rieure, Universit\'{e} de Recherche Paris Sciences et Lettres, 75005 Paris, France}
\author{Jesse Ault}
\affiliation{School of Engineering, Brown University, Providence, RI 02912, USA}
\author{Thomas Salez}
\affiliation{Univ. Bordeaux, CNRS, LOMA, UMR 5798, F-33405, Talence, France}
\affiliation{Global Station for Soft Matter, Global Institution for Collaborative Research and Education, Hokkaido University, Sapporo, Hokkaido 060-0808, Japan}
\author{Clarissa Sch\"{o}necker}
\affiliation{Technische Universit\"{a}t Kaiserslautern, 67663 Kaiserslautern, Germany}
\affiliation{Max Planck Institute for Polymer Research, 55218 Mainz, Germany}
\author{Howard A. Stone}
\affiliation{Department of Mechanical and Aerospace Engineering, Princeton University, Princeton, New Jersey 08544, USA}
\author{L. Mahadevan}
\email{lmahadev@g.harvard.edu}
\affiliation{School of Engineering and Applied Sciences, Department of Physics, Kavli Institute for Nano-Bio Science and Technology, Harvard University, Cambridge, MA 02138, USA}
\begin{abstract}
A submerged finite cylinder moving under its own weight along a soft incline lifts off and slides at a steady velocity while also spinning.  Here, we experimentally quantify the steady spinning of the cylinder and show theoretically that it is due to a combination of an elastohydrodynamic torque generated by flow in the variable gap,  and the viscous friction on the edges of the finite-length cylinder. The relative influence of the latter depends on the aspect ratio of the cylinder as well as the deformability of the substrate, which we express in term of a single scaled compliance parameter. By varying this compliance parameter, we show that our experimental results are consistent with a transition from an edge-effect dominated regime for short cylinders to a gap-dominated elastohydrodynamic regime when the cylinder is very long. 
\end{abstract}
\maketitle

\section{Introduction}

The interplay between lubricated flow and deformable surfaces is ubiquitous in nature and engineering in settings spanning a broad range of length scales, \textit{e.g.} earthquakes \cite{Ma2003}, avalanches \cite{Glenne1987}, landslides \cite{Campbell1989}, lubrication of cartilaginous and artificial joints \cite{Maroudas76, Greene2011, Grodzinsky1978, Mow1984, Mow2002, Bouchet2015} or industrial bearings \cite{Hamrock1994}. Often, this elastohydrodynamic coupling is seen in the presence of confined flow where pressure gradients are likely to be large. Previous theoretical works have studied confined flows in the soft lubrication approximation and accounted for the roles of elasticity \cite{sek93,sko04,sko05,sno13,Beaucourt2004,Urzay2007}, fluid compressibility \cite{Balmforth2010}, the inertia of the fluid and the elastic medium \cite{Clarke2011}, and viscoelasticity of the substrate \cite{Pandey2016}. More recent works have focused on elastohydrodynamic effects for liquids confined at the micro and nano scales \cite{Villey2013, Karan2018, Zhang2019}, which has important consequences for surface mechanical characterization \cite{Leroy2012, Wang2017}. For symmetrical objects, the results show that elastic deformations lead to a non-symmetric pressure field and to the emergence of a friction-reducing lift force. Of particular importance in nature are cases of freely moving particles close to soft surfaces as seen in flows of cells in vessels \cite{Goldsmith1971} or microfluidic devices \cite{Byun2013,Davies2018}, the mobility of suspended or falling objects near elastic membranes \cite{Daddi2016, Daddi2017, Ral18, Daddi2018}, the behavior of vesicles near walls \cite{Abkarian2002} or the collisions between suspended particles \cite{Davis1986}. It is only very recently that a theoretical work \cite{sal15} addressed freely moving objects and showed how a free falling cylinder can sediment, slide and spin along a soft incline. A particularly interesting result is that the elastohydrodynamic lift force can counteract sedimentation and lead to an emergent sliding steady state that has since been confirmed experimentally \cite{sai16}.  The experimental study also raised a new question associated with observations of rotational motion, which led to a recent theoretical study of the rotation \cite{ral2017} that remains untested.

         In this article, we experimentally quantify the rotation of cylinders falling along a soft incline. We show that there is a steady rotation speed for finite-length cylinders that increases with substrate deformability, qualitatively consistent with a recently developed theory for an infinite cylinder near a soft substrate \cite{ral2017}. However, the latter fails to describe quantitatively our results. We show that a complete theory that takes into account both the elastohydrodynamic torque along the cylinder length and the viscous friction on the edges of the cylinder is in quantitative agreement with our experiments and that our control parameters can be combined into a single dimensionless compliance number. When this compliance increases, \textit{i.e.}, the thickness of the substrate increases or its stiffness decreases, the angular velocity follows a relationship that contains two regimes, a first one dominated by edge effects and the second by the elastohydrodynamic stresses due to the substrate deformation. In contrast with the theory for infinite cylinders developed previously, here the edge effects do not allow for the existence of simple power law behaviors in the range of our experimental parameters.
         
\section{Experimental system and observations}

\begin{figure}[h!]
	\centering
	\includegraphics[scale=0.3]{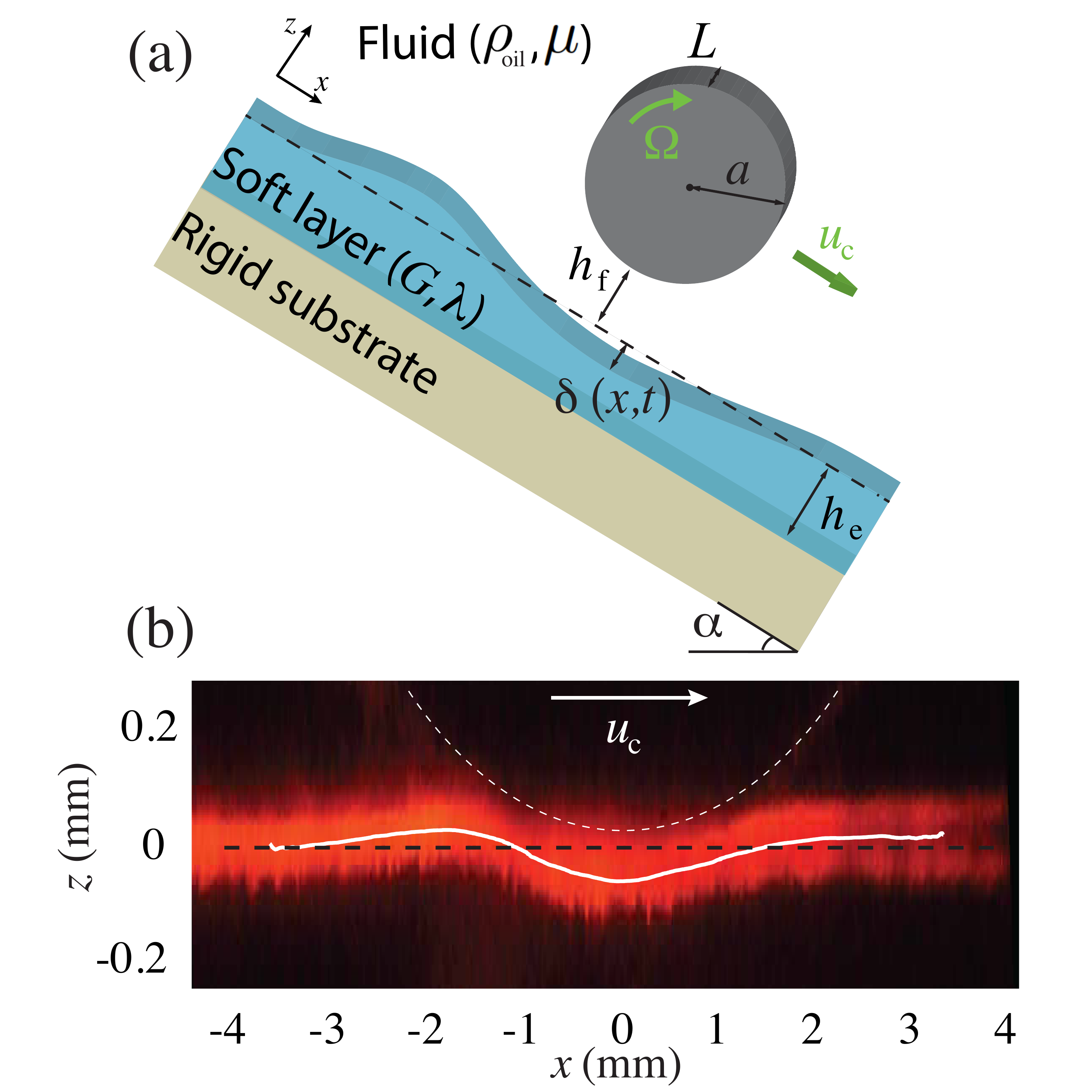}
	\caption{(a) Sketch of the experimental setup:  a negatively-buoyant rigid cylinder immersed in a viscous bath slides down a tilted wall that is coated with a thin elastic layer. (b) Experimental image showing a side view of the soft substrate deformation (red) by using a laser sheet with fluorescent particles placed at the surface. The white dashed line represents the cylinder contour, centered at $x = 0$. The black dashed line corresponds to the interface of the undeformed substrate. The white solid line follows the center of the fluorescent particles' emission, obtained by using a Gaussian fit, showing the asymmetric deformation of the substrate-fluid interface. The experimental parameters are $G=65$ kPa, $h_\textrm{e}=1.5$ mm, $\mu=1$ Pa.s, $a=12.7$ mm, $\rho=8510$ kg/m$^3$, and $\alpha=11^\circ$. Figure adapted and modified from \cite{sai16}.}
	\label{fig:setup}
\end{figure}

The experiments follow the same protocol as described previously \cite{sai16}, with a metal cylinder of either aluminum or brass (densities $\rho=2720$ and $8510$ kg/m$^3$) with radii  $a=12.7$ and $6.35$ mm. For both cylinders, the length $L=12.7$ mm such that their respective aspect ratios are $a/L=1$ and $1/2$.  The cylinders are immersed in a silicone oil bath of density $\rho_\textrm{oil}=970 $ kg/m$^3$ and viscosity $\mu=[0.35-30]$ Pa.s. They freely move down a rigid glass incline (angle varied in the range $\alpha=[11-45]^\circ$) coated with a soft gel with shear moduli $G$ in the range $[100 - 3\times10^5]$ Pa (Fig. \ref{fig:setup}(a)). The coating thickness is varied in the range $h_\textrm{e}=[100-2000]~\mu$m. The coatings are made of polydimethylsiloxane (PDMS) and polyacrylamide (PAA) in which we can change the concentrations of monomers and crosslinkers to tune the  shear modulus. The latter is measured on an Anton Paar MCR501 rheometer with a CP50 cone-plate geometry, using an amplitude of $0.1\%$ for PAA and $0.5\%$ for PDMS, with an angular frequency of $10$ rad/s. When the cylinder moves along the incline (undergoing both translation and rotation), it deforms the substrate (Fig. \ref{fig:setup}(b)) and its motion is recorded from the side with a camera. 

\begin{figure}[h!]
	\centering
	\includegraphics[scale=0.45]{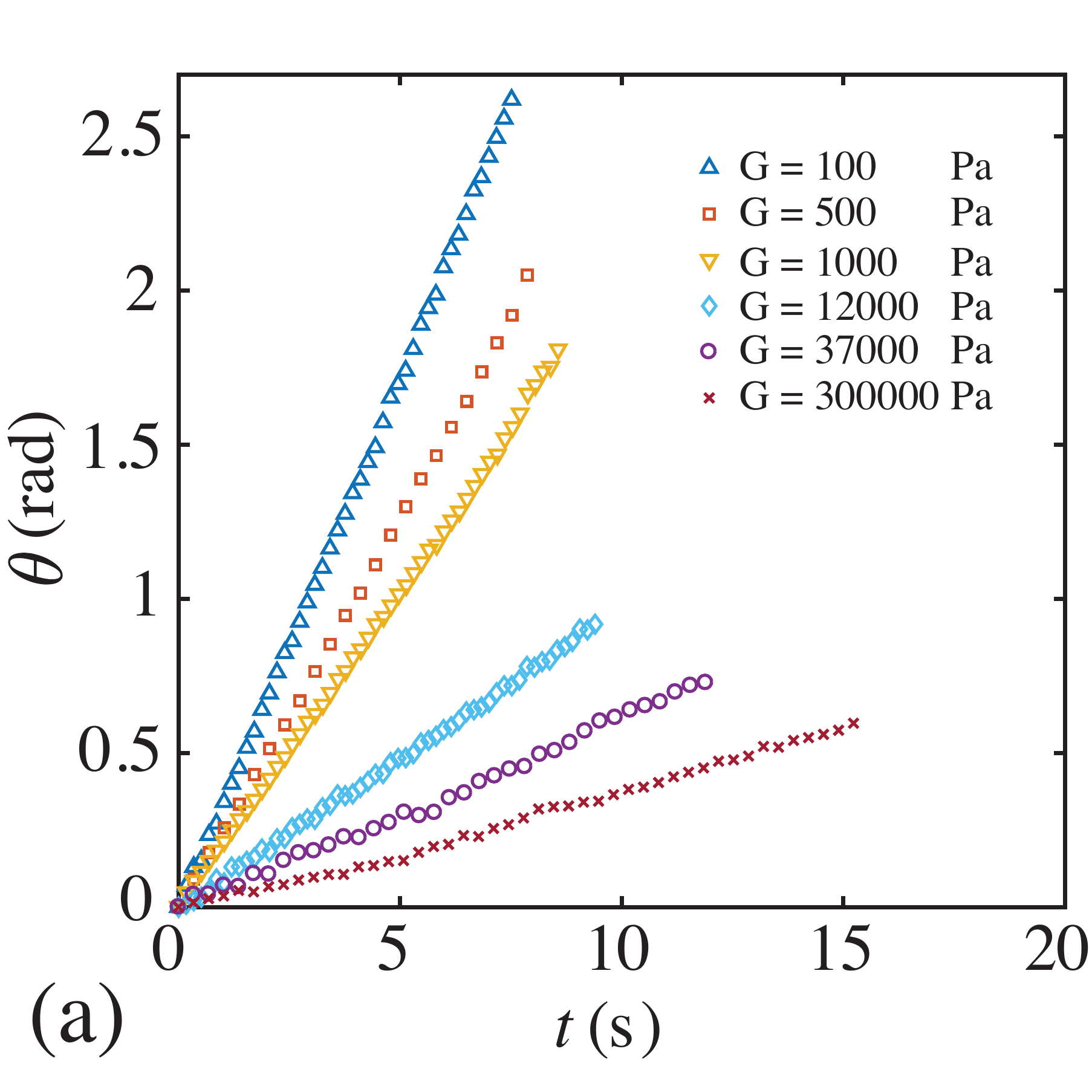}
	\includegraphics[scale=0.45]{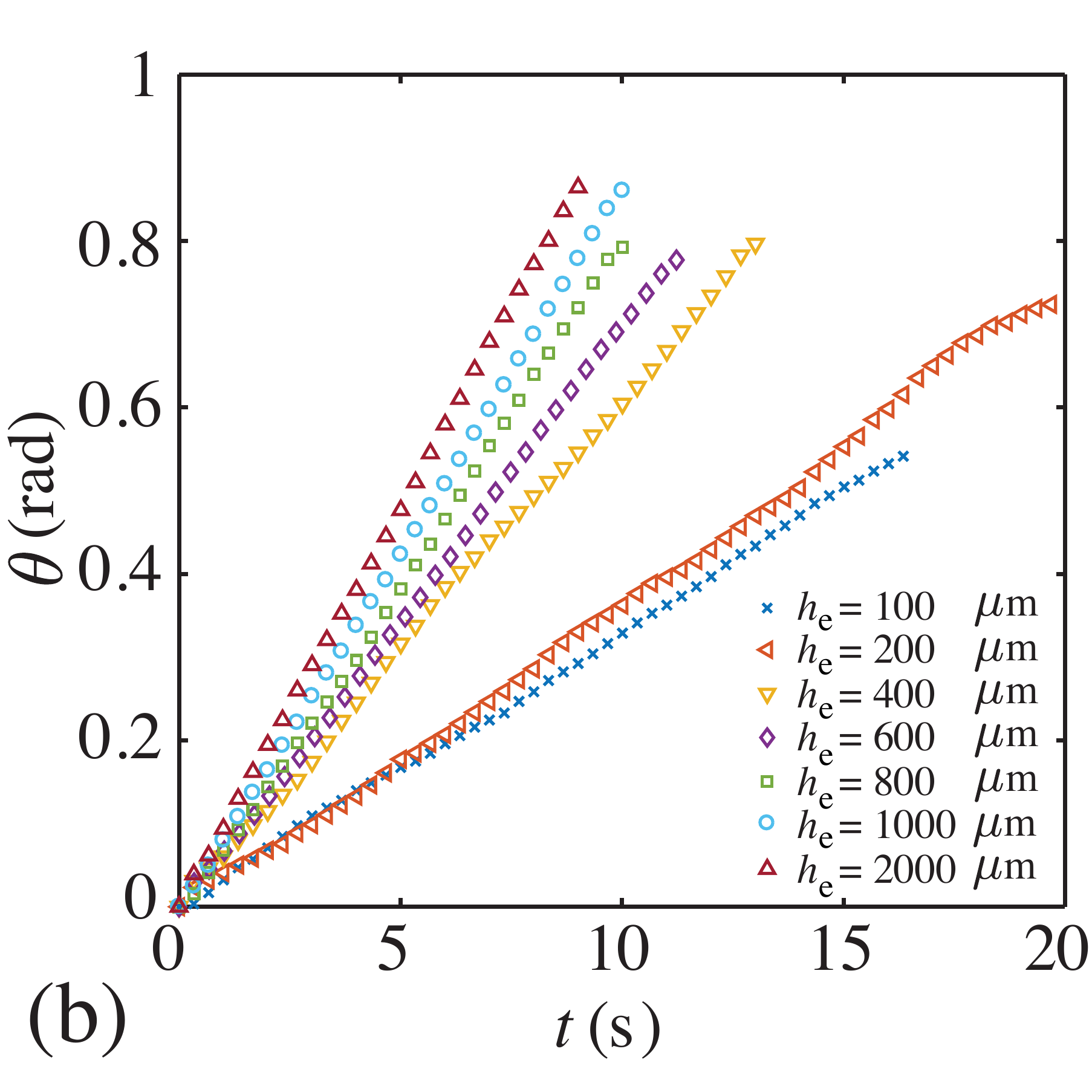}
	\caption{(a) Evolution of the rotation angle of the cylinder as a function of time for different shear moduli of the coating with the aluminium cylinder of radius $a=12.7$ mm. These experiments were conducted at constant coating thickness $h_\textrm{e}=600~\mu$m. (b) Evolution of the rotation angle as a function of time for different coating thicknesses with the aluminum cylinder.  These experiments were made at constant shear modulus $G=31\times10^3$ Pa. For both panels, the viscosity and the incline angle are fixed at $\mu=1$ Pa.s and $\alpha=11^\circ$ respectively.}
	\label{fig:alpha(t)}
\end{figure}

         Image analysis allows us to track the center of the cylinder and provides a direct measurement of the translation speed of the cylinder $u_\textrm{c}$, and its rotation velocity $u_{\theta}=a\Omega$, with $a$ the cylinder radius and $\Omega$ the angular speed. Figures \ref{fig:alpha(t)}(a) and (b) show the rotation angle as a function of time for the aluminum cylinder for different coating moduli and thicknesses, respectively. We observe that the  rotation speed $\Omega$ is constant, which is reminiscent of the constant sliding speed observed earlier in similar experiments \cite{sai16}. We also observe that this rotation speed decreases when the coating becomes less deformable, \textrm{i.e}, when the shear modulus $G$ increases, or when its thickness $h$ decreases. 

\section{Scaling arguments and the finite size effect}

To capture the main features of the experimental observations, we first revisit the scaling arguments for an immersed infinite cylinder sliding along a soft incline \cite{sal15}. Due to the confinement of the flow under the cylinder within a fluid gap of thickness $h_\textrm{f} \ll a$, the typical transverse length scale of contact  scales as $\ell = \sqrt{2 a h_\textrm{f}}$, so that we can invoke the lubrication approximation \cite{Reynolds1886}. For small deformations $\delta$ of the soft layer satisfying  $|\delta|\ll \ell$ and assuming a localized linear response of the elastic layer to the applied normal stress $p$ (Winkler approximation), the deformation can be expressed as $\delta = \frac{h_\textrm{e}}{2G+\lambda}p$, where $\lambda$ denotes Lam\'{e}'s first parameter of the substrate and where the hydrodynamic pressure in the gap scales from the Stokes equation as $p \sim \mu u_\textrm{c} \ell/h_\textrm{f}^2$.  This identifies the dimensionless compliance of the elastic layer $\Lambda$ as
\begin{equation}\label{lambdaDef}
    \Lambda \equiv \frac{\mu u_\textrm{c} h_\textrm{e} a^{1/2}}{(2G + \lambda) h_\textrm{f}^{5/2}} \sim \frac{\delta}{h_\textrm{f}},
\end{equation}
which measures the scale of the substrate deformation relative to the fluid gap thickness \footnote{The definition of $\Lambda$ used here differs from the definition in \citep{ral2017} by a factor of $\sqrt{2}$.}. In this framework, previous theoretical studies \cite{sek93,sko04,sko05,sno13,Beaucourt2004,Urzay2007} have shown that for $\Lambda \ll 1$ and for given $u_\textrm{c}$ and $h_\textrm{f}$ the motion of an infinite cylinder is accompanied by the emergence of an elastohydrodynamic lift force $F \sim \Lambda \mu u_\textrm{c} \ell^2 L/h_\textrm{f}^2$, which was confirmed experimentally \cite{sai16}. Since the cylinder also rotates with negligible inertia, the sum of torques due to elastohydrodynamics (induced by the substrate's deformation due to sliding) $\tau_\textrm{s}$ and viscous damping of the rotational motion $\tau_{\Omega}$ vanish  \cite{ral2017}:  $\tau_{\Omega}+\tau_\textrm{s}=0$. The sliding torque scales as  $\tau_\textrm{s} \sim \mu u_\textrm{c} a \ell L/(h_\textrm{f}+\delta)$, where $h_\textrm{f}+\delta$ is the typical gap size between the cylinder and the deformed substrate. Invoking \eqref{lambdaDef}, we expand the previous expression in powers of $\Lambda$ for $\Lambda \ll 1$ and recognize that the contributions proportional to $\Lambda^0$ and $\Lambda^1$ are zero for an \emph{infinite} cylinder \cite{jef81,sal15} to find that $\tau_\textrm{s} \sim \mu u_\textrm{c} a \ell \delta^2 L/h_\textrm{f}^3$. As the rotational damping torque scales as $\tau_{\Omega} \sim \mu \Omega a^2 \ell L/h_\textrm{f}$, balancing it with the sliding contribution yields the scaling relationship:
\begin{equation}
    \frac{a \Omega}{u_\textrm{c}}\sim\Lambda^2\sim\frac{ \mu^2 u_\textrm{c}^2 h_\textrm{e}^2 a}{(2 G + \lambda)^2 h_\textrm{f}^{5}} \quad \mbox{(infinite cylinder)}.
    \label{u_theta_ad}
\end{equation}
For infinite cylinders, no rotation occurs when $\Lambda = 0$ (a rigid substrate), a consequence of a vanishing sliding torque in this limit \cite{jef81}. This feature is modified for compact bodies such as spheres, where translation and rotation are coupled even when all boundaries are rigid.

\begin{figure}
	\centering
	\includegraphics[scale=0.4]{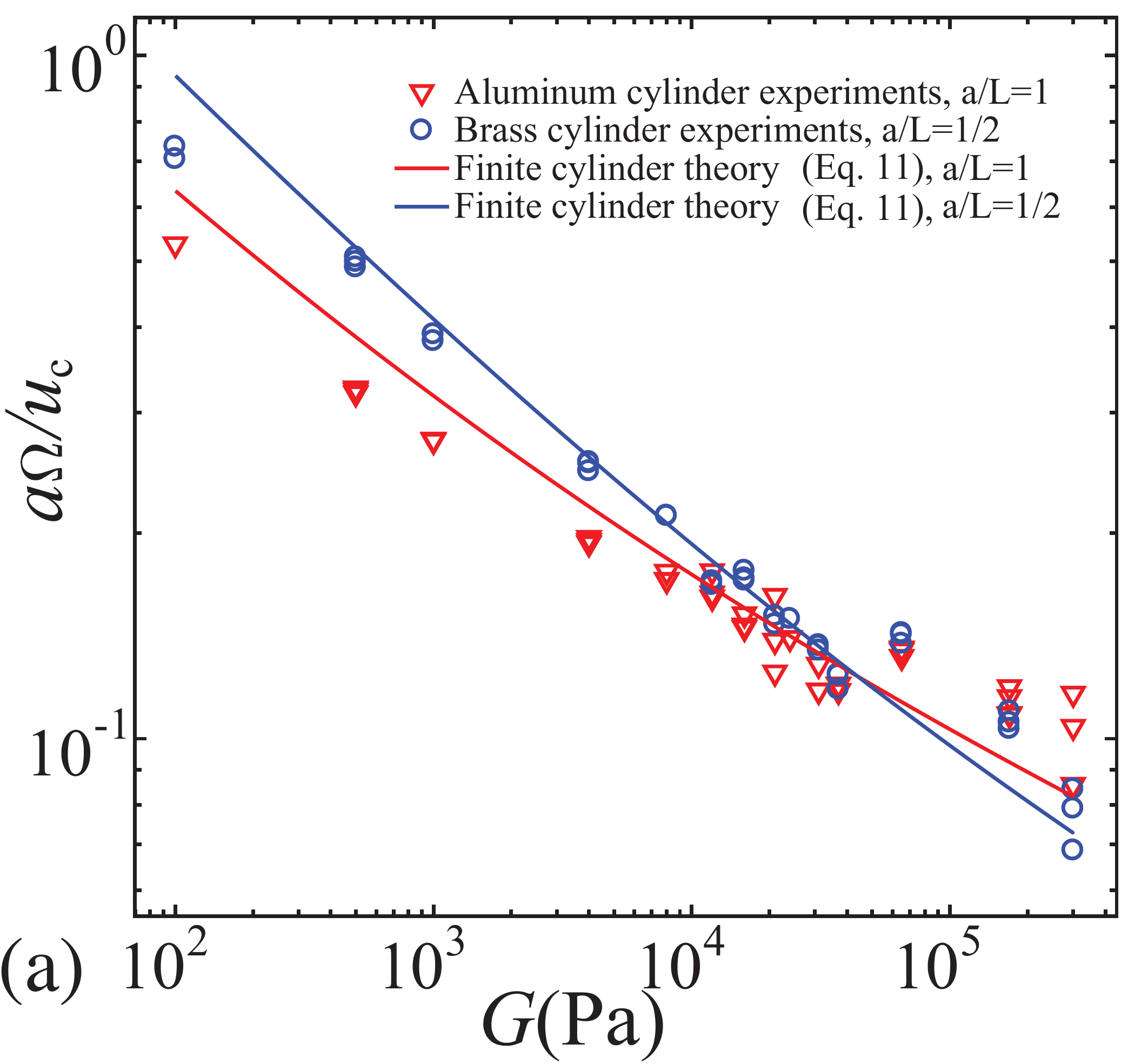}
	\includegraphics[scale=0.4]{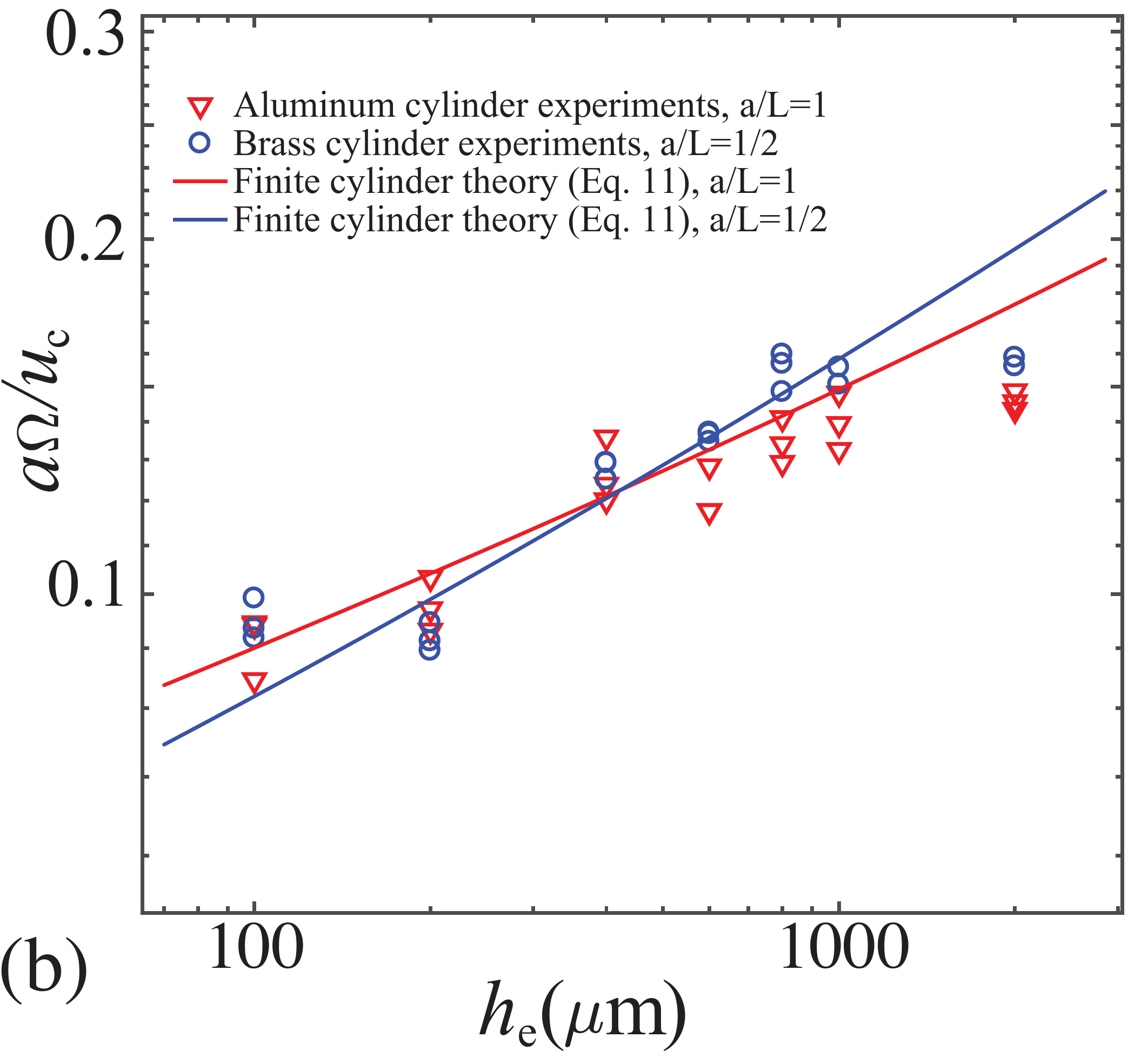}
	\caption{(a) Dimensionless angular velocity $a\Omega/u_\textrm{c}$ as a function of the elastic substrate's shear modulus $G$, for the aluminum cylinder of radius $a=12.7$ mm ($a/L=1$), and the brass cylinder, of radius $a=6.35$ mm ($a/L=1/2$). The thickness of the elastic substrate is $h_\textrm{e}=600~\mu$m. The solid lines correspond to the theoretical prediction of \eqref{RotationTheory}  for $a/L=1$ (red) and $a/L=1/2$ (blue) with $c=-0.715$. (b) Dimensionless angular velocity $a\Omega/u_\textrm{c}$ as a function of the elastic substrate's thickness $h_\textrm{e}$, for the aluminum cylinder of radius $a=12.7$ mm ($a/L=1$), and the brass cylinder of radius $a=6.35$ mm ($a/L=1/2$). The shear modulus of the elastic substrate is $G=31\times10^3$ Pa. The solid lines correspond to the theoretical prediction \eqref{RotationTheory} for $a/L=1$ (red) and $a/L=1/2$ (blue) with $c=-0.715$ as a fit parameter. For both studies, the viscosity and incline angle are $\mu=1$ Pa.s and $\alpha=11^\circ$ respectively. The standard deviation obtained from the angle measurements as a function of time, and averaged on all the experimental points, is $0.05$.}
	\label{fig:Vratio}
\end{figure}

For the finite-sized cylinders in experiments, we generically expect a nonzero rotation rate $\Omega_0(h_\textrm{f}/a, a/L)$ even as $\Lambda \rightarrow 0$ due to three-dimensional flows near the cylinder ends. These flows penetrate a width $\ell$ into the fluid gap from the ends of the cylinder. A rough estimate of the resulting sliding torque (for a rigid substrate) is $\mu u_\textrm{c} a \ell ^2 /h_\textrm{f} \sim \mu u_\textrm{c} a^2$, which is independent of $h_\textrm{f}$. In lubrication flows with gap-independent scaling estimates for torque, detailed calculations typically reveal logarithmic corrections \citep{gol67a, Claeys1988, Kim_Karrila}. Including a log-corrected end torque $\mu u_\textrm{c} a^2 \log (a/h_\textrm{f})$ in the torque balance above suggests
\begin{equation} \label{TwoTermScaling}
    \frac{a \Omega}{u_\textrm{c}} = k_2 \Lambda^2+k_1\frac{a}{L} \left(\frac{h_\textrm{f}}{a}\right)^{1/2} \log \left(\frac{a}{h_\textrm{f}}\right).
\end{equation}
The second term on the right side is identified with $a \Omega_0/u_c$ and the first with \eqref{u_theta_ad}, with constants of proportionality $k_1$ and $k_2$. Thus, we expect two independent sources of rotation, one due to end effects and another due to the elastohydrodynamic torque over the length of the cylinder. As we will show, there is indeed a cross-over from  end-dominated to softness-dominated rotation  in our experiments as $\Lambda$ increases. 

\section{Theory}
As discussed above, two-dimensional theory predicts zero hydrodynamic torque on a non-rotating (infinite) cylinder sliding along rigid walls ($\Lambda = 0$). We show below that three-dimensional end-effects qualitatively modify this result for a cylinder. End effects are confined to a penetration depth $\ell$ into the lubrication gap, so both ends are hydrodynamically isolated in our experiments since $L = O(a) \gg \ell$. We focus on the flow near one of the ends, which we place at $y = 0$ so that the gap lies in $y > 0$. It is convenient to introduce dimensionless coordinates $(X, Y) = (x, y)/\ell$, and a dimensionless lubrication pressure $P(X,Y) = p(x,y)/(\mu u_\textrm{c} \ell/h_\textrm{f}^2)$. Since the gap thickness abruptly diverges at the ends of the cylinder, $P$ must vanish at $Y = 0$. Invoking the parabolic approximation of the gap profile $H(X) = 1 + X^2$ and focusing on $\Lambda \rightarrow 0$ (the limit of a rigid substrate), the pressure in the gap satisfies the Reynolds equation
\begin{subequations}\label{LubricationPDE}
\begin{align}
    &\nabla \cdot (H^3 \nabla P + 6 H \ex) = 0\,, \qmbox{subject to} \\
    &P(X,0) = \pp{P}{Y}(X,\infty) = P(\pm \infty,Y) = 0,
\end{align}
\end{subequations}
where $\nabla = \ex \ppi{}{X} + \ey \ppi{}{Y}$.

We seek a solution $P(X,Y) = P_{\rm 2d}(X) + P'(X,Y)$, where $P_{\rm 2d} (X)= 2X/(1+X^2)^2$ is the pressure due to an sliding infinite cylinder, which satisfies \eqref{LubricationPDE} except for the condition at $Y = 0$. As we discuss below, it is sufficient to analyze the large-$X$ behavior of $P'$. Defining $\eta = Y/X$ (the tangent of the angle in the $XY$ plane), we seek an asymptotic solution in inverse powers of $X$ with the form $P'(X \gg 1,Y) \sim \sum_{n} X^{-n} f_n(\eta)$. From the boundary condition at $Y = 0$ and the asymptotic behavior $P_{\rm 2d}(X \gg 1) \sim 2 X^{-3}$, it is clear that the leading term of the expansion introduced above is $P'(X \gg 1,Y) \sim - 2 X^{-3} Q(\eta)$. Substituting this expression into (\ref{LubricationPDE}a) and retaining the most slowly decaying terms at large $X$ yields
\begin{subequations}
\begin{align}
&(1 + \eta^2) \dd[2]{Q}{\eta} + 2 \eta \dd{Q}{\eta}  - 6 Q = 0, \qmbox{subject to} \\
&Q(0) = 1 \qmbox{and} \dd{Q}{\eta}\bigg|_{\eta \rightarrow \infty} \rightarrow 0,
\end{align}
\end{subequations}
which admits the solution
\begin{align}
Q(\eta) = \left(3 \eta ^2+1\right) \left(1 - \frac{2}{\pi} \arctan \eta\right)- \frac{6 \eta}{\pi}.
\end{align}
This determines the asymptotic behavior $P'(X \gg 1,Y) \sim -2 X^{-3} Q(Y/X)$. The perturbation scheme can developed further to obtain corrections to $P'$ [the next term is of the form $X^{-5} f_5(\eta)$] although the leading term suffices for our purposes.  

 The dimensionless horizontal velocity in the reference frame of the sliding cylinder, expressed in units of $u_c$ is $\boldsymbol{\mathbf{V}} = \frac{1}{2} Z(Z-H) \nabla P + \frac{Z - H}{H} \ex$. The component of the shear stress responsible for its rotation, in units of $\mu u_c/h$, is $\sigma_{XZ} = \pp{V_X}{Z}\big|_{Z = H} = \frac{H}{2} \pp{P}{X} + \frac{1}{H}$, whose integral over the area of the lubrication gap yields the hydrodynamic sliding torque on the cylinder. Noting the symmetry of $\sigma_{XZ}$ about $X = 0$, including both (hydrodynamically non-interacting) ends of the cylinder, and recalling that the torque generated by the two-dimensional case is identically zero, the dimensionless torque can be expressed (in units of $\mu u_{\textrm{c}} a \ell^2/h_\textrm{f}$) as $4  \int_0^{X_{\infty}} \int_{0}^{\infty} \frac{H}{2} \pp{P'}{X} \,\rmd Y \rmd X$, where $X_{\infty} = O(a/\ell)$ [corresponding to $x = O(a)$] is the outer ``edge'' of the lubrication gap. An estimate of the previous integral at large $X$ shows that  it diverges as $\log X_{\infty}$. Formally, we make a change of variables in the integral from $(X,Y)$ to $(X, \eta)$ and isolate the divergence to obtain the dimensional sliding torque 
\begin{align} \label{TorqueEnds}
    \tau_\textrm{c} &= \frac{4\mu u_\textrm{c} a \ell^2}{h_\textrm{f}} \int^{X_{\infty}} \int_0^{\infty} \frac{1}{X} \left(3 Q + \eta \dd{Q}{\eta} \right) \rmd \eta \,\rmd X\nonumber \\
    &= \frac{32}{3 \pi}\mu u_\textrm{c} a^2 \left(\log \left(\frac{a}{h_\textrm{f}}\right) + c\right).
\end{align}
The constant $c$ absorbs the ambiguity in defining $X_{\infty}$, nonsingular contributions from the lubrication flow (\textit{i.e.} from terms of $P'$ decaying as $X^{-5}$ or faster) and the torque due to end-effects outside the fluid gap. The latter contribution includes the torque on the flat faces of the cylinder, which is generated by stresses of $O(\mu u_c/a)$ acting over an area of $O(a^2)$ with a moment arm of $O(a)$. Evaluating $c$ requires a matched asymptotic approach that we do not pursue here; instead we will estimate it from a fit to our experiments. The result \eqref{TorqueEnds} is reminiscent of the torque on a translating sphere of radius $a$, for which the factor of $32/(3 \pi)$ is replaced by $4 \pi/5$ and the constant $c \approx -1.895$ \citep{gol67a}.

Since the cylinder is free to rotate and has negligible inertia, the sum of the sliding torque and the rotational torque $\tau_{\Omega} = -2 \sqrt{2} \pi \mu a^2 L \Omega_0 (a/h_\textrm{f})^{1/2}$ \citep{jef81} vanishes,  yielding the rotation rate of a translating finite cylinder near a rigid wall 
\begin{equation} \label{RotationEndEffets}
    \frac{a \Omega_0}{u_\textrm{c}} = \frac{8 \sqrt{2}}{3 \pi^2} \frac{a}{L} \left(\frac{h_\textrm{f}}{a}\right)^{1/2} \left(\log\left(\frac{a}{h_\textrm{f}}\right)  + c\right).
\end{equation}
This result is expected to dominate for stiff substrates $(\Lambda \ll 1)$ in our experiments.  The leading contribution to $\Omega$ due to the softness of the substrate (denoted $\Omega_2$)  was shown for an infinite cylinder to be $ a \Omega_2/u_\textrm{c} = (21/128) \Lambda^2$ \citep{ral2017}. Modifications to $\Omega_2$ due to end effects scale as $\ell/L \ll 1$ and will be neglected here. 

Thus, the angular speed of a translating finite cylinder is $\Omega \approx \Omega_0 + \Omega_2$, or
\begin{equation} \label{RotationTheory}
    \frac{a \Omega}{u_\textrm{c}} = \frac{8 \sqrt{2}}{3 \pi^2} \frac{a}{L} \left(\frac{h_\textrm{f}}{a}\right)^{1/2}\! \left(\log\left(\frac{a}{h_\textrm{f}}\right)  + c\right)  + \frac{21}{128} \Lambda^2.
\end{equation}
 This theoretical prediction makes precise the estimate \eqref{TwoTermScaling} and reduces to the infinite-cylinder and the rigid-wall results in the respective limits $a/L \rightarrow 0$ and $\Lambda \rightarrow 0$.

For gravity-driven motion along a soft incline, the gap thickness is not an independently controlled quantity but is instead set by a balance of the cylinder's buoyant weight, the elastohydrodynamic lift force and the hydrodynamic drag on the cylinder. Introducing the Poisson ratio $\nu$ [so that $\lambda = 2 G \nu/(1 - 2 \nu)$], this balance yields for a thin, compressible, elastic layer \cite{sal15,ral2017}
\begin{subequations} \label{GravitySliding}
\begin{align}
\frac{h_\textrm{f}}{a} &= \left(\frac{3}{8} \Lambda \tan \alpha \right)^2 \qmbox{with} \\
\Lambda &= \left\{\frac{2^{21/10}}{3^{4/5}} \left(\frac{1 - 2\nu}{1 - \nu}\right)^{1/5}\right\} \kappa 
\end{align}
\end{subequations}
where $\kappa=\left(\frac{\rho^* g h_\textrm{e} \cos \alpha}{2 G \tan^3 \alpha}\right)^{1/5}$ and $\rho^*=\rho-\rho_{oil}$. In (\ref{GravitySliding}b), the quantity in braces is a dimensionless constant with values $1.78$--$0.93$ for $\nu$ in the range $0$--$0.49$, while all parameters involved in $\kappa$ are either known or directly measured. Substituting (\ref{GravitySliding}a) into \eqref{RotationTheory} yields the angular speed for gravity-driven motion near a thin, compressible coating on an incline of angle $\alpha$;
\begin{align} \label{RotationTheory2}
    \frac{a \Omega}{u_\textrm{c}} &= \frac{\sqrt{2}\, a}{\pi^2 L} \left(\Lambda \tan \alpha\right) \left(2 \log\left(\frac{8}{3 \Lambda \tan \alpha}  \right)  + c\right) \nonumber \\ &\quad + \frac{21}{128} \Lambda^2.
\end{align}
End effects dominate the rotation rate at small $\Lambda$, although the gap thickness is still set by elastohydrodynamic stresses. The term quadratic in $\Lambda$ becomes important when $\Lambda \gtrsim (a/L) \tan \alpha$. In the limit of very stiff substrates, we expect $a \Omega/u_\textrm{c} \propto \left(h_\textrm{e}/G\right)^{1/5} \log \left(G/h_\textrm{e}\right)$, in contrast with the two-dimensional prediction $a \Omega/u_\textrm{c} \propto \left(h_\textrm{e}/G\right)^{2/5}$.

\begin{figure}[!h]
	\centering
	\includegraphics[scale=0.3]{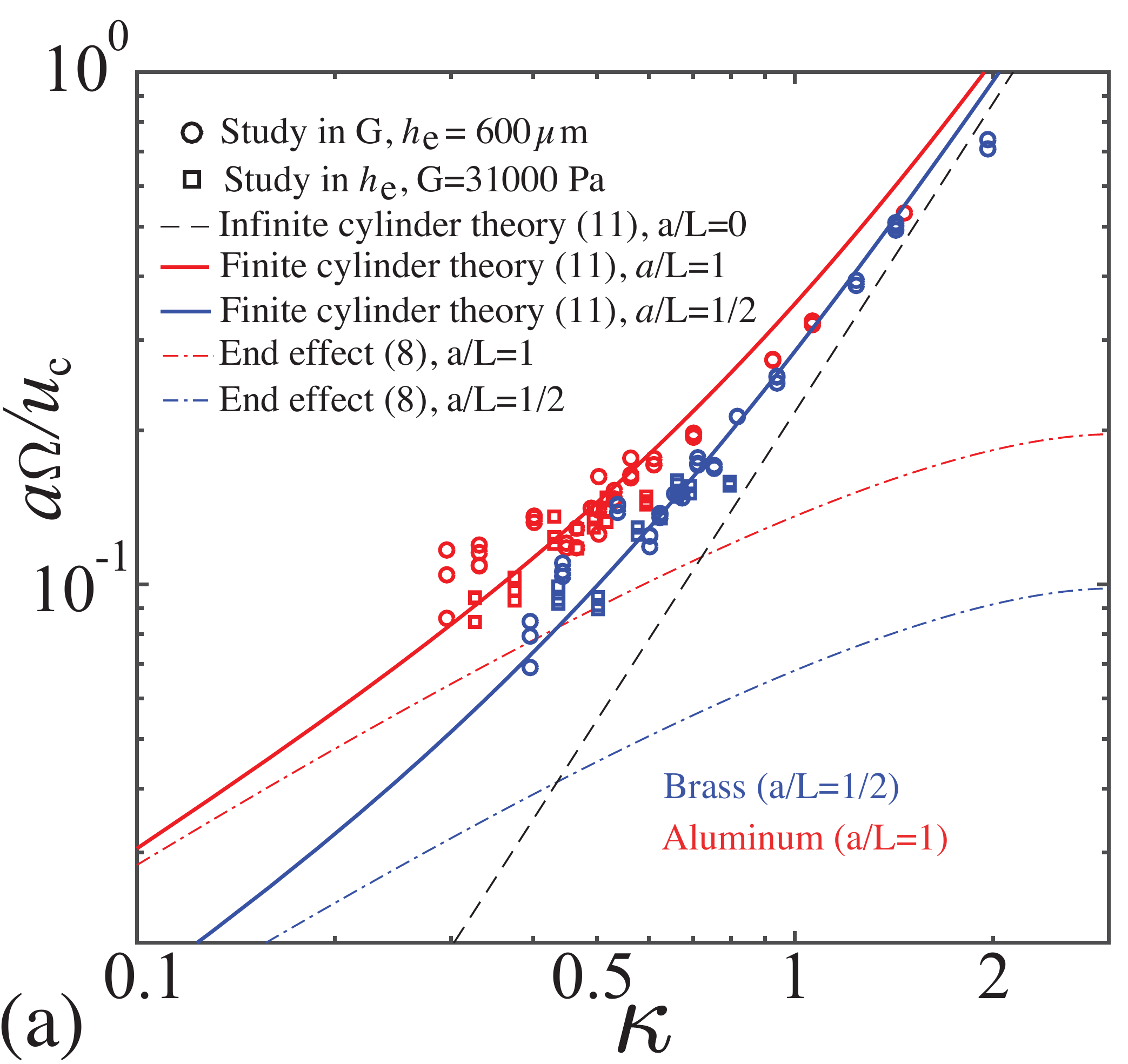}
	\includegraphics[scale=0.3]{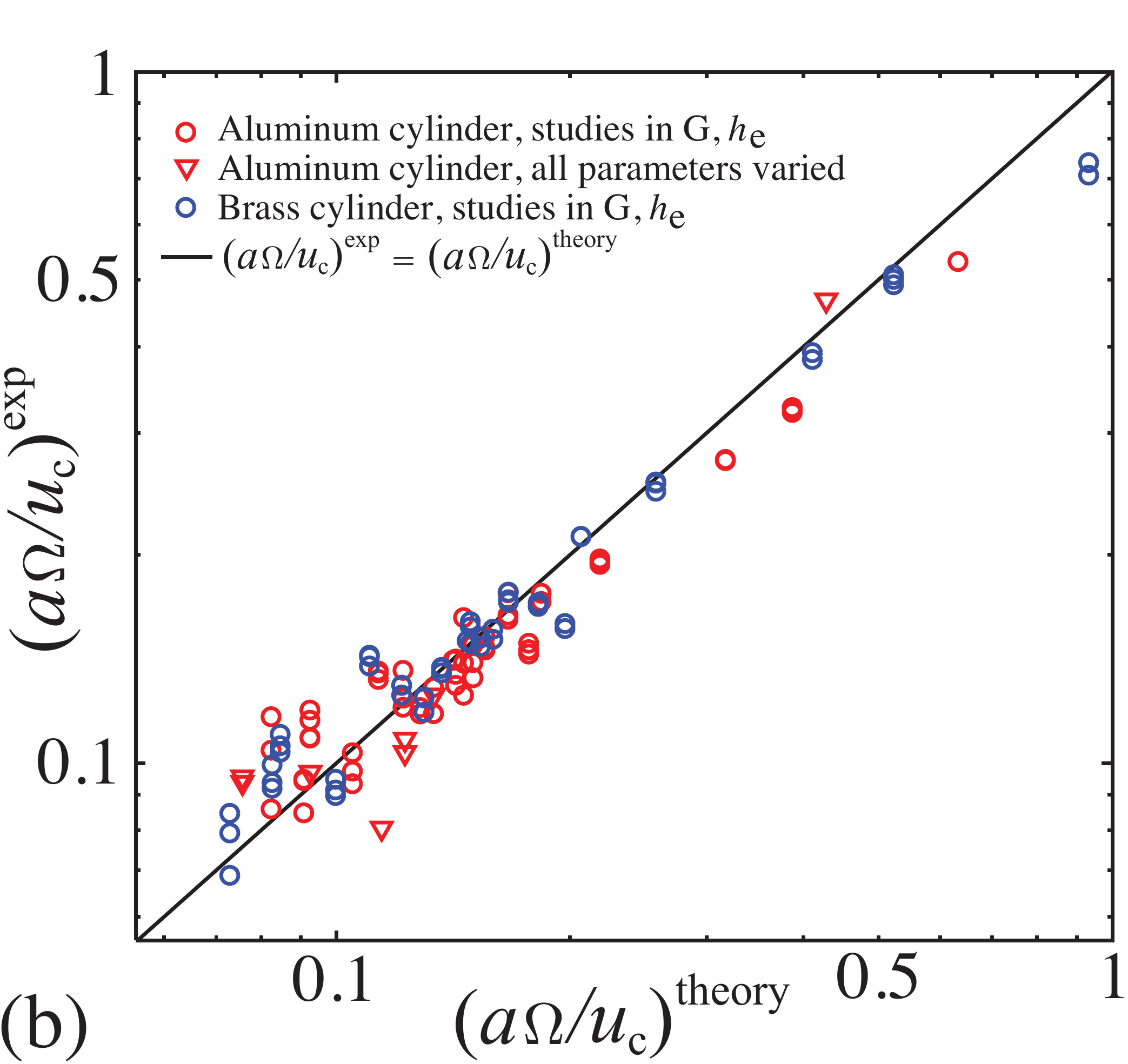}
	\caption{(a) Experimental dimensionless angular velocity $a\Omega/u_\textrm{c}$ as a function of the modified scaled compliance $\kappa=\left(\frac{\rho^* g h_\textrm{e} \cos \alpha}{2 G \tan^3 \alpha}\right)^{1/5}$. The red symbols correspond to the aluminum cylinder with $a/L = 1$, while the blue symbols correspond to the brass cylinder with $a/L = 1/2$. The circles and the squares correspond, respectively, to variations in $G$ and $h_\textrm{e}$. The black dashed line corresponds to the infinite cylinder case \eqref{RotationTheory} with $a/L=0$, $\nu=0.47$. The colored dashed lines correspond to the theory taking into account only the end effect \eqref{RotationEndEffets}, with $c=-0.715$. The solid lines corresponds to the finite-size theory \eqref{RotationTheory}, with $c=-0.715$. (b) Experimental scaled angular velocity $(a\Omega/u_\textrm{c})^{\textrm{exp}}$ as a function of the theoretical scaled angular velocity $(a\Omega/u_\textrm{c})^{\textrm{theory}}$ \eqref{RotationTheory}. Red circles - aluminum cylinder with $a/L = 1$, blue circles - brass cylinder with $a/L = 1/2$,  triangles - aluminum cylinder ($a/L = 1$) with viscosities $\mu \in [0.35-30]$ Pa.s, incline angles $\alpha \in [11-45^\circ]$, moduli $G \in [16 -280]$ kPa, thicknesses $h_\textrm{e} \in [300 -1000]~\mu$m, and the black line has slope 1.}
	\label{fig:Master}
\end{figure}

\section{Comparison between experiments and theory}

We now compare the prediction of the theory with the results of the experiments. The evaluation of the compliance $\Lambda$ in \eqref{lambdaDef} and \eqref{GravitySliding}(a) requires us to know the value of the Poisson ratio $\nu$. Since we did not measure it directly, we choose $\nu=0.47$ based on experimental studies found in the literature for polyacrylamide gels and thin films of PDMS \cite{Takigawa1996,Boudou2009,Dogru2018}.  The theoretical prediction for the scaled rotational speed $a\Omega/u_\textrm{c}$ in \eqref{RotationTheory} includes a constant $c$ that is expected to be independent of the compliance $\Lambda$. As $\ell/L \ll 1$ we assume the end flows to be decoupled from each other and we thus expect $c$ to be independent as well of the aspect ratio of the cylinder. In order to compare the theory to the experiments we force $c$ to be the same for experiments involving different cylinders (and thus aspect ratios). In Fig. \ref{fig:Vratio}, we show the behavior of the scaled angular speed $a\Omega/u_\textrm{c}$ as a function of the coating film's shear modulus $G$ (Fig. \ref{fig:Vratio}a) and thickness $h_\textrm{e}$ (Fig. \ref{fig:Vratio}b). We observe that the finite-size theory, which includes both the cylinder edge-effect term and an elastohydrodynamic term (the latter corresponding to an infinite soft-lubricated cylinder) predicts remarkably well the experimental results with a single constant $c=-0.715$, with increasing scaled angular velocities for decreasing stiffness $G$ and increasing coating thickness $h_\textrm{e}$ (increasing $\kappa$). The value for $c$ is consistent with the typical value obtained for a sphere near a rigid wall ($c \approx -1.895$ \citep{gol67a}).

Combining all these experimental results allows us to plot a master curve for $a \Omega /u_\textrm{c}$ as a function of the modified scaled compliance $\kappa=\left(\frac{\rho^* g h_\textrm{e} \cos \alpha}{2 G \tan^3 \alpha}\right)^{1/5}$, as shown in Fig. \ref{fig:Master}(a). We choose to plot the data as a function of $\kappa$ rather than $\Lambda$ as the former only accounts for the experimental parameters that we can directly measure, and not $\nu$ or $h_\textrm{f}$. The values of $\kappa$ are very similar to those of $\Lambda$ for $\nu=0.47$ ($\Lambda \approx 1.15 \kappa$). In fact the factor between $\Lambda$ and $\kappa$ is rather insensitive to $\nu$ (\textit{e.g.} about $0.93$ for $\nu = 0.49$), and so $\kappa$ is a good physical estimate of the scaled compliance $\Lambda$ for our experimental conditions. We observe that, with a unique constant $c=-0.715$, the experimental results are very consistent with the theoretical master curves. In Fig.  \ref{fig:Master}(b), we have plotted the values measured for $a \Omega /u_\textrm{c}$ as a function of its theoretical prediction from \eqref{RotationTheory}, for the same $c$ constant and the same data as in Fig. \ref{fig:Master}(a), but also with experiments where all parameters were varied, including the inclination angle. This unique master curve for both cylinders confirms the good agreement between theory and experiments over more than a decade.

\section{Discussion}

We have also plotted separately the contributions of both terms in \eqref{RotationTheory}, namely the contribution of end effects for a finite-length cylinder, and the elastohydrodynamic contribution for an infinite cylinder, as shown in Fig. \ref{fig:Master}(a). Our experimental data lie in the crossover region between these two limiting behaviors. At high values of the compliance \textit{i.e.}, for soft or thick substrates, the experimental data for both aspect ratios appear to collapse together and converge toward the infinite cylinder theory, consistent with a regime where edge effects (and thus cylinder length) do not affect the rotation behavior. We note that at intermediate values of the compliance, edges effects tend to increase the scaled angular velocity with respect to the infinite-cylinder prediction. Finally, at small compliances, the elastohydrodynamic torque does not affect the rotation anymore, and the latter is solely generated by end effects (near a rigid wall). The crossover location depends on the aspect ratio. We can indeed see that, for the brass cylinder with $a/L=1/2$, the rotation behavior is closer to the infinite cylinder one than in the case of the aluminum cylinder, with $a/L=1$, where end effects play a more significant role. 
     
It is also interesting to note that the theory predicts an angular velocity either smaller or larger than in $a \Omega /u_\textrm{c}\sim1$. The latter regime corresponds to the rolling of a cylinder in no-slip dry contact with a rigid incline and should be reached in our system typically for $\kappa\sim2$. However, the range of parameters explored in our experiments could not allow us to verify the existence of ``super-rolling'' behaviors for higher compliances.

\section{Conclusion}

Our experiments on the rotation of an immersed finite-size cylinder moving down and near a soft incline have shown that there is a steady-state rotation with an angular speed that increases with the compliance of the substrate. While this observation is qualitatively consistent with a recent theoretical prediction for an infinite cylinder \cite{ral2017}, this earlier infinite cylinder (2D) theory fails to describe our experimental observations quantitatively. A modified theoretical description for a finite-length cylinder that takes into account the additional torque created by viscous friction on both its edges does allow for a quantitative agreement  with our experiments, which are typical of many applications. In particular, we have shown that for small compliances and small cylinder lengths, the contribution of the elastohydrodynamic torque to the rotation becomes small relative to those contributions from end effects, even when the gap thickness is still set by a finite elastohydrodynamic lift force. This result gives more realistic insights on the behaviors of finite-size objects in motion or in interaction close to soft interfaces, and pave the way for new theoretical development accounting for geometric and mechanical properties that are relevant to more specific biological, geophysical and engineering processes.

\acknowledgments
HAS acknowledges support from the National Science Foundation via award CMMI-1661672. CS  acknowledges support from the German Research Foundation (DFG) - Project-ID 172116086-SFB 926.


\end{document}